\documentclass[fleqn,10pt]{wlscirep}

\usepackage{color}
\usepackage{graphicx}
\usepackage{amsmath}
\usepackage{hyperref}
\usepackage{latexsym}

\newcommand{\beq}{\begin{eqnarray}}
\newcommand{\eeq}{\end{eqnarray}}

 \def\>{\rangle}

\title{Creating a switchable optical cavity with controllable quantum-state mapping between two modes}

\author[1,*]{Grzegorz Chimczak}
\author[1,2]{Karol Bartkiewicz}
\author[3,4]{Zbigniew Ficek}
\author[1]{Ryszard Tanaś}

\affil[1]{Faculty of Physics, Adam Mickiewicz University, PL-61-614 Pozna\'n, Poland}
\affil[2]{RCPTM, Joint Laboratory of Optics of Palack\'y
University and Institute of Physics of Academy of Sciences of the
Czech Republic, 17. listopadu 12, 772 07 Olomouc, Czech Republic}
\affil[3]{The National Centre for Applied Physics, KACST, P.O. Box 6086, Riyadh 11442, Saudi Arabia}
\affil[4]{Quantum Optics and Engineering Division, Institute of Physics, University of Zielona G\'ora, Szafrana 4a, Zielona G\'ora 65-516, Poland}

\affil[*]{chimczak@amu.edu.pl}

\begin{abstract}
We describe how an ensemble of four-level atoms in the diamond-type configuration can be applied
to create a fully controllable effective coupling between two cavity modes. 
The diamond-type configuration allows one to use a bimodal cavity that supports
modes of different frequencies or different circular polarisations, because 
each mode is coupled only to its own transition.
This system can be used for mapping a quantum state of one cavity mode onto the other mode on demand.
Additionally, it can serve as a fast opening high-{\cal{Q}} cavity system that can be easily and
coherently controlled with laser fields.
\end{abstract}
\begin{document}

\flushbottom
\maketitle

\thispagestyle{empty}

\section*{Introduction}
Quantum systems, in which a control of the coherent evolution is possible, are of great
importance from a theoretical and a practical points of view, and therefore, such systems
always attract research interest~\cite{bo17,liu17}. Systems composed of a cavity and atoms,
which are trapped inside this cavity, are such systems because one can easily control
the evolution of their quantum state just by illuminating atoms with
a laser~\cite{McKeeverSCI04,boozerPRL07,weberPRL09,nollekePRL13,hackerNAT16}.
Moreover atom-cavity systems provide a versatile environment for engineering complex
non-classical states of light~\cite{larsonJOM06,pradoPRA11,domokosEPJD98,savage90,rongChPB11,
nikoghosyanPRL12,liuQIP13,ficekPRA13,xiaoQIP13,liuQIP14}.
Researchers achieve such high level of control over the evolution of quantum states
employing atoms, which can be modelled by few special level schemes. 
The simplest and frequently considered schemes are three level atoms in $\Lambda$ and $V$
configurations~\cite{imamoglu99,miranowicz02,boozer07,harkonen09,jiong12,zhang14,casabone15,dong16}.
The main advantage of these atoms is the possibility of working with the two-photon Raman transition
involving an intermediate level, which is populated only virtually during the whole evolution.
Since atoms are driven by a classical laser field, the Raman transition takes place
only if the laser is turned on. The same idea allows for full control of the system evolution
in many other level schemes. Therefore researchers have used and studied intensively many different
types of atoms coupled to the cavity mode~\cite{lin08,xiao08,sharypov13,everitt14,yavuz05,serra05,
xiao06,kurz14,lloyd01,clark03,wilkSCI07}. There is, however, one important atomic level scheme, which
is almost ignored by researchers in the context of atom-cavity systems --- a four-level atom
in the diamond configuration (a $\Diamond$-type atom, also known as a double-ladder four-level atom).
Despite the fact that this level scheme is rich in quantum interference and coherence features~\cite{Ou2009}
and has many other applications~\cite{yan06,kajari07,becerra08,stowe08,hsu11,kolle12,lee13,parniak15},
to the best of our knowledge there are only few articles about the $\Diamond$-type atom coupled to
the quantized field modes~\cite{grynberg90,rathe95,hu99,deng08,qamar09,li2011,qiang12,ge13,wang13,baghshahi14,offer16,baghshahi16}.

\begin{figure}[ht]
\centering
 \includegraphics[width=6.5cm]{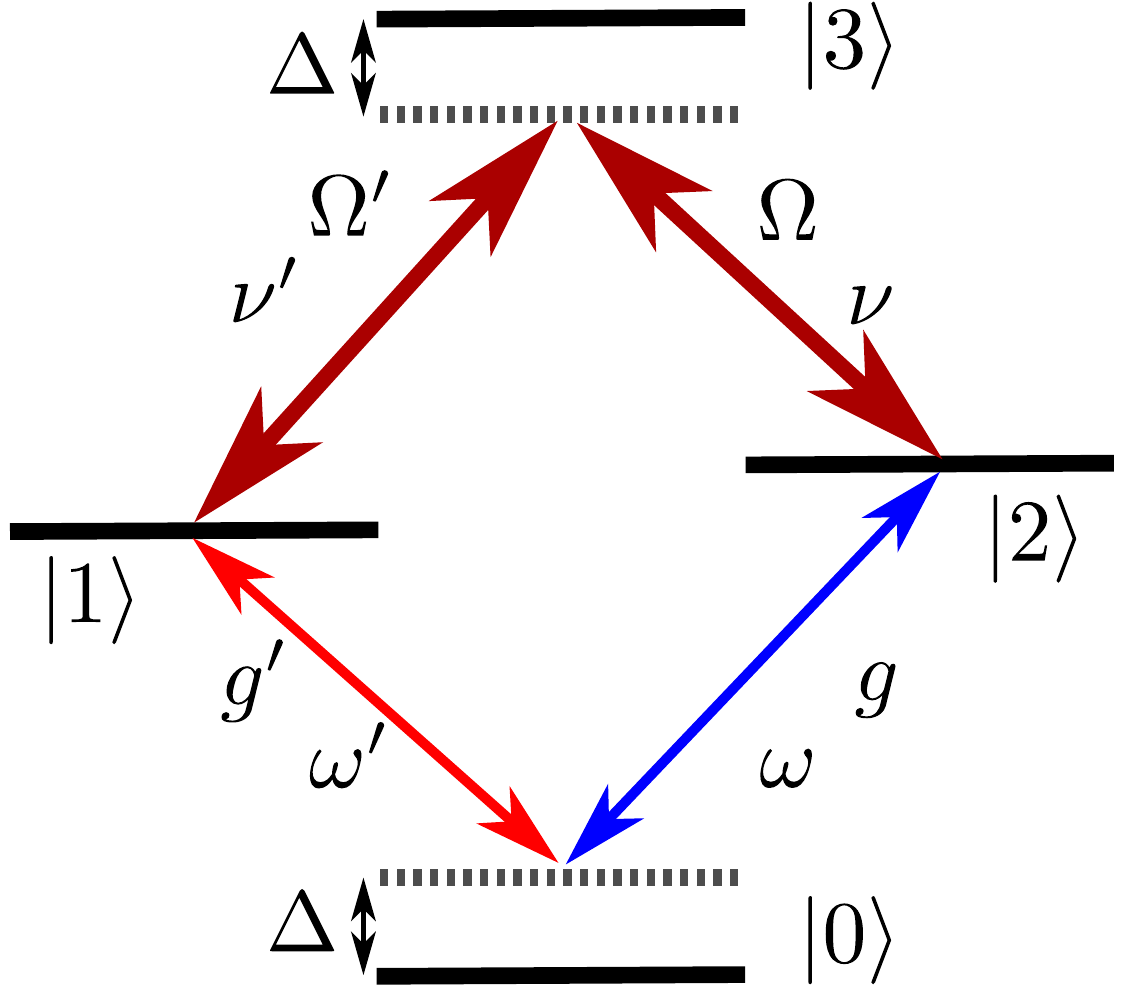}
  \caption{  
Energy levels of an atom in the diamond configuration. Lower atomic transitions are
coupled to quantized field modes with frequencies $\omega$ and $\omega'$. Upper
transitions are driven by classical laser fields with frequencies $\nu$ and $\nu'$.}
\label{fig:diamond}
\end{figure}

In this paper we study a $\Diamond$-type atom interacting with two quantized cavity modes
and two classical laser fields. 
The quantized field modes are coupled to lower atomic transitions while the classical 
laser fields are coupled to upper atomic transitions, as depicted in figure~\ref{fig:diamond}.
Here, we show that under certain conditions
the evolution of this system can be described by a simple effective Hamiltonian, and can be
easily controlled just by switching the lasers on and off. We also present two applications of this
system. First of them is the transfer of an arbitrary state of light from one mode to the other.
Second application is a device that plays the role of an effective cavity, in which we can change
the effective {\cal{Q}} factor on demand just by turning the lasers on and off. This device is based
on the scheme proposed by Tufarelli~\emph{et~al.}~\cite{tufarelliPRL14} but it employs $\Diamond$-type
atoms instead of two-level atoms which makes the physics of the described 
system much richer. Thus, the proposed system  has the potential to be more versatile and efficient in quantum information processing than the solutions based on the two-level atoms.

%------------------------------------------------------------------
\section*{Results}\label{sec:model}
\vspace*{5mm}
\noindent
\textbf{Effective description of the system.}
We consider an ensemble of $n$ identical four-level atoms in the diamond configuration 
(figure~\ref{fig:diamond}) with a ground level $|0\rangle$, 
two non-degenerate intermediate levels $|1\rangle$, $|2\rangle$, and an upper level $|3\rangle$. 
There are four allowed transitions in this level scheme. The $|0\rangle\leftrightarrow|2\rangle$
transition is coupled to the field mode represented by the annihilation operator $a$ with 
coupling strength $g$, while the $|0\rangle\leftrightarrow|1\rangle$ transition is coupled to 
the field mode described by the annihilation operator $b$ with coupling strength $g'$. 
The frequency of the $a$ mode is $\omega$ and the frequency of the $b$ mode is $\omega'$.
Both field modes are equally detuned from the corresponding transition frequencies
by $\Delta=(E_1-E_0)/\hbar-\omega'=(E_2-E_0)/\hbar-\omega$. 
The upper transitions
$|1\rangle\leftrightarrow|3\rangle$ and $|2\rangle\leftrightarrow|3\rangle$
are driven by coherent laser fields of frequencies $\nu'$ and $\nu$, respectively.
The coupling strengths between these atomic transitions and
the laser fields are denoted by $\Omega'$ and $\Omega$. Both laser
fields are detuned from the corresponding transition frequencies by
$\Delta$. Simultaneously, the atom is coupled to all other modes of the EM field,
which are assumed to be in the vacuum state. 
The atom provides an effective coupling between both the modes. Of course,
the effective coupling strength depends on the number of atoms $n$.
The higher the number of atoms $n$, the stronger the coupling becomes.
We assume that there are $n\ge 1$ identical $\Diamond$-type
four-level atoms trapped inside the cavity.
The evolution of this composite quantum system is governed by the Hamiltonian,
which in the rotating frame is given by
\begin{equation}
  \label{eq:Hamiltonian0}
  H=\sum_{k=1}^{n}\big\{\Delta \sigma_{11}^{(k)}+\Delta \sigma_{22}^{(k)}+
  2\Delta \sigma_{33}^{(k)}+(\Omega\sigma_{23}^{(k)}
+\Omega'\sigma_{13}^{(k)}
  +g a^{\dagger} \sigma_{02}^{(k)}+g' b^{\dagger} \sigma_{01}^{(k)}
  + {\rm{h.c.}})\big\}\, ,
\end{equation}
where $\hbar=1$ and $\sigma_{ij}^{(k)}=|i\rangle_{k}\langle j|$
denotes the atomic flip operator between states $|i\rangle_k$ and $|j\rangle_k$
for the $k$th atom.
The Lindblad operators representing spontaneous transitions from the atomic
excited states are given by
\begin{eqnarray}
  \label{eq:L1}
  L_{1}^{(k)}=\sqrt{\gamma'}\sigma_{01}^{(k)}\, , \quad 
  L_{2}^{(k)}=\sqrt{\gamma}\sigma_{02}^{(k)}\, ,\quad
  L_{3}^{(k)}=\sqrt{\gamma_{3}}\sigma_{23}^{(k)}\,, \quad 
  L_{4}^{(k)}=\sqrt{\gamma'_{3}}\sigma_{13}^{(k)}\, ,  
\end{eqnarray}
where $\gamma$, $\gamma'$, $\gamma_{3}$ and $\gamma'_{3}$ are spontaneous
emission rates for the respective transitions. 
For the sake of simplicity, we assume 
that $\Omega$, $\Omega'$, $g$ and $g'$ are real, non-negative numbers.
Similar four-level scheme has been proposed in Ref.~[\citenum{sharypov13}].
The diamond configuration, however, has the advantage that it allows to
use (contrary to level scheme of Ref.~[\citenum{sharypov13}]) atomic transitions with
the highest values of the dipole moment. Of course, the higher the dipole moment,
the stronger the effective coupling between the modes is.
Using the method of adiabatic elimination (see Methods) 
we derive the following effective Hamiltonian
\begin{eqnarray}
  \label{eq:HamiltonianEFF}
  H_{\rm{eff}}&=&\delta_0 (a^{\dagger}a+b^{\dagger}b)+
  \delta_1 b^{\dagger}b+\delta_2 (a^{\dagger}b+b^{\dagger}a)\, ,
\end{eqnarray}
where $\delta_0=-n g^2 \alpha_2$, $\delta_1=n (g^2\alpha_2-g^{\prime 2}\alpha_1)$
and $\delta_2=-n g g' \alpha_3$,  for $\alpha_1=\xi (\Omega^2-2\Delta^2)$,
$\alpha_2=\xi (\Omega^{\prime 2}-2\Delta^2)$,
$\alpha_3=-\xi\Omega\Omega'$, $\alpha_4=-\xi\Delta^2$, $\alpha_5=\xi\Delta\Omega'$,
$\alpha_6=\xi\Delta\Omega$ with 
$\xi=1/(\Delta[\Omega^2+\Omega^{\prime 2}-2\Delta^2])$.
This effective Hamiltonian (\ref{eq:HamiltonianEFF}) 
works properly if populations of all atomic excited states 
are small (see Alexanian-Bose method in Methods). 
The effective master equation
\begin{equation}
  \label{eq:masterEFF}
  \dot{\rho}=-i [H_{\rm{eff}},\rho]
  +\sum_{j=1}^{2} \left\lbrace L_{\rm{eff}}^{(j)}\rho (L_{\rm{eff}}^{(j)})^{\dagger}
  -\frac{1}{2} \left[(L_{\rm{eff}}^{(j)})^{\dagger}L_{\rm{eff}}^{(j)}\rho
  +\rho (L_{\rm{eff}}^{(j)})^{\dagger}L_{\rm{eff}}^{(j)}\right]\right\rbrace \, ,
\end{equation}
where
\begin{eqnarray}
  \label{eq:Leff_2}
  L_{\rm{eff}}^{(1)}=\sqrt{n \gamma'}
  [\alpha_3 g a+\alpha_1 g' b]\, , \quad
  L_{\rm{eff}}^{(2)}=\sqrt{n \gamma}
  [\alpha_2 g a+\alpha_3 g' b]\, ,
\end{eqnarray}
requires more restrictive conditions to work properly, because in its derivation 
(see Reiter-S{\o}rensen method in Methods) we have neglected the Lindblad operators
$L_3^{(k)}$ and $L_4^{(k)}$, which describe spontaneous emissions from
the upper states $|3\rangle_{k}$. Therefore, we assume that populations
of the atomic intermediate levels ($|1\rangle_{k}$ and $|2\rangle_{k}$) are small 
and populations of the upper states $|3\rangle_{k}$ are small even compared with 
the intermediate levels, because then probabilities of occurrence of collapses
described by $L_{3}^{(k)}$ and $L_{4}^{(k)}$ are negligibly small. It is necessary
to know conditions for the parameters, which make these assumptions true. We restrict
ourselves only to cases where $\Omega' g\approx\Omega g'$. In these cases the 
effective master equation works properly if the following conditions are satisfied
\begin{eqnarray}
  \label{eq:condA}
&|\Delta|\gg g_{\rm{min}}\sqrt{n}\,\max\left(\sqrt{\langle a^{\dagger}a\rangle},\sqrt{\langle b^{\dagger}b\rangle}\right)\quad {\rm{and}}\quad 
\max(\lambda_2, \lambda_3, \lambda_5, \lambda_6)\ll\min(\lambda_1, \lambda_4)\, ,&
\end{eqnarray}
where $g_{\rm{min}}=\min(g,g')$ and  $\lambda_i$ are dimensionless expansion parameters
(see Alexanian-Bose method in Methods for the definition of $\lambda_i$ parameters). 
The expansion parameters $\lambda_2$, $\lambda_3$, $\lambda_5$ and $\lambda_6$ are
associated with operators acting on the states $|3\rangle_{k}$. The smaller they are,
the smaller are the populations of the states $|3\rangle_{k}$. The expansion 
parameters $\lambda_1$ and $\lambda_4$ are associated with operators acting 
only on the states $|1\rangle_{k}$ and $|2\rangle_{k}$). Knowing values of $g$ and $g'$ 
of the chosen physical system we set the value of $\Delta$ according to the first
condition and then we find numerically the value of $\Omega$ for which
the second condition is satisfied. We can always find such value of $\Omega$, because
when intensities of classical fields tend to infinity, then expansion parameters
$\lambda_2$, $\lambda_3$, $\lambda_5$, $\lambda_6$ tend to zero.
In the following text we are going to use the effective master equation~(\ref{eq:masterEFF}).

%\subsection{The limit of high-intensity classical fields}
Let us consider the dynamics of the four-level atom in the diamond configuration in
the limit of high-intensity classical fields. In the dressed-state approach 
there is one ground atomic state $|0\rangle$ and three excited states 
\begin{eqnarray}
  \label{eq:HamiltonianA3}
  |\mu\rangle&=&{\cal{N}}_\mu
  (-\Omega |1\rangle+\Omega' |2\rangle)\, ,\nonumber\\
  |\phi\rangle&=&{\cal{N}}_\phi
  (2\Omega'|1\rangle+2\Omega |2\rangle+(\Delta-\Omega_{\rm{R}})|3\rangle)
  \, ,\nonumber\\
  |\psi\rangle&=&{\cal{N}}_\psi 
  (2\Omega'|1\rangle+2\Omega |2\rangle+(\Delta+\Omega_{\rm{R}})|3\rangle)\, ,  
\end{eqnarray}
where ${\cal{N}}_\mu$, ${\cal{N}}_\phi$ and ${\cal{N}}_\psi$ are normalisation
factors and $\Omega_{\rm{R}}=(\Delta^2+4\Omega^2+4\Omega^{\prime 2})^{1/2}$.
Here, there are three allowed transitions: $|0\rangle\leftrightarrow|\mu\rangle$,
$|0\rangle\leftrightarrow|\phi\rangle$ and $|0\rangle\leftrightarrow|\psi\rangle$,
each of which is coupled to both cavity modes (see Alexian-Bose method in Methods).
As mentioned above, when intensities of classical fields tend to infinity,
then expansion parameters $\lambda_2$, $\lambda_3$, $\lambda_5$, $\lambda_6$
tend to zero. It means that only two atomic levels, \emph{i.e.} $|0\rangle$ and 
$|\mu\rangle$, are enough to describe the evolution of the system --- the four-level
atom in the diamond configuration effectively works exactly in the same way as
the detuned two-level atom in this regime. Note that the excited bare state
$|3\rangle$ can be then neglected. It might seem counter-intuitive
that high coupling strengths between the atomic upper transitions and the laser fields
lead to an effective decoupling of the upper level $|3\rangle$ from the system dynamics,
but this idea is known and discussed for example in Ref.~[\citenum{kyoseva12}].

In the limit of high-intensity classical fields one more thing is clearly seen
from the first condition in~(\ref{eq:condA}) --- the effective coupling strength 
$\delta_{2}$ scales as $\sqrt{n}$. Such behaviour is the well known feature of
the collective dynamics~\cite{beige05,kim18,ferraro18}.

%\subsection{The effective Hamiltonian for lasers turned off}
When the lasers are turned off then the evolution of the system is still governed
by the Hamiltonian~(\ref{eq:Hamiltonian0}) but with $\Omega=\Omega'=0$.
The formulas for the effective Hamiltonian given by equation~(\ref{eq:HamiltonianEFF})
and the effective operators in this case read as
\begin{eqnarray}
  \label{eq:EFF0}
  H_{\rm{eff}}=-(n g^2/\Delta)\, a^{\dagger}a-(n g^{\prime 2}/\Delta)\, b^{\dagger}b\, ,\quad
  L_{\rm{eff}}^{(1)}=\sqrt{n \gamma'} (g'/\Delta)\, b\, ,\quad  L_{\rm{eff}}^{(2)}=\sqrt{n \gamma} (g/\Delta)\, a\, .
\end{eqnarray}
Here, we can also easily derive more precise expressions, if we perform the adiabatic elimination of excited atomic states assuming from the start that $\Omega=\Omega'=0$. This approach results in
\begin{eqnarray}
  \label{eq:Hamiltonian_closed}
  H_{\rm{eff}}=-\frac{n g^2\Delta}{\Delta^2+\gamma^2/4} a^{\dagger}a
  -\frac{n g^{\prime 2}\Delta}{\Delta^2+\gamma^{\prime 2}/4} b^{\dagger}b\, ,\quad
  L_{\rm{eff}}^{(1)}=\frac{\sqrt{n \gamma'}g'}{\Delta-i\gamma'/2} b\, ,\quad
  L_{\rm{eff}}^{(2)}=\frac{\sqrt{n \gamma}g}{\Delta-i\gamma/2} a\, .
\end{eqnarray}
It is important to note that there is no coupling between the two cavity modes,
and therefore, there is no photon transfer when the lasers are turned off.
We will refer to this working mode of the system as to {\emph{the closed mode}}.

%------------------------------------------------------------------

%------------------------------------------------------------------
%\section*{Quantum-state mapping between two cavity modes}\label{sec:mapping}
\vspace*{5mm}
\noindent
\textbf{Quantum-state mapping between two cavity modes.}
Under certain conditions the evolution of a complex system formed by an ensemble of
four-level diamond-type atoms interacting with two quantized field modes
can be easily controlled just by switching the lasers on and off. 
Let us now demonstrate that we can use this system to transfer a given
quantum state of one mode (for example a qudit or the Schr\"odinger's cat states) to
the other mode on demand. It has shown that in special cases, \emph{i.e.}, for coherent states
and for qubit states, the Hamiltonian of the form~(\ref{eq:HamiltonianEFF}) can swap
the states of the two modes~\cite{lin08,xiao08}.
Here, we show that it is possible to transfer an arbitrary photonic state.

\begin{figure}[ht]
\centering
  \includegraphics[width=6.5cm]{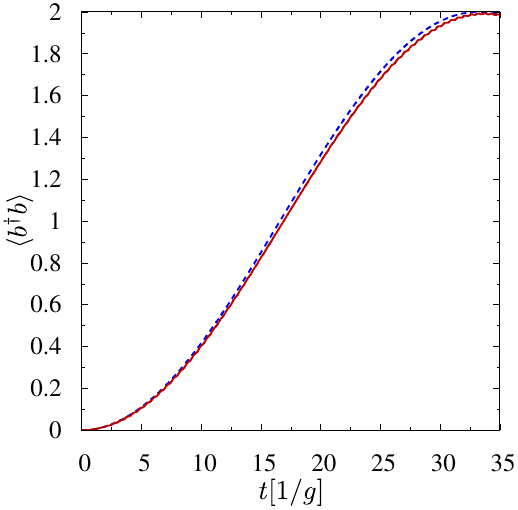}
   \caption{The average photon number in the $b$ mode as a function of time calculated numerically (solid line) and
   given by equation~(\ref{eq:mapping01}) (dashed line) for one atom $n=1$ and
   $(g',\Delta,\Omega,\Omega',\gamma,\gamma',\gamma'')/g=(1,11,55,55,1,1,1)$, where $g/2\pi=10$~MHz.
   At $t=0$ the $a$ mode is prepared in the state $|2\rangle_{A}$, while the $b$ mode is in a vacuum state. After the $\pi$
   pulse both photons are transferred to the $b$ mode.}
  \label{fig:btb}
\end{figure}

First, we need the formula for the average photon number
in the mode represented by the annihilation operator $b$, assuming that initially this
mode is empty, while the mode represented by $a$ is prepared in the Fock state $|n_{\rm{ph}}\rangle$.
This formula will help us investigate the photon transfer process. We can derive it
introducing the superposition bosonic operator of both field modes
\begin{eqnarray}
  \label{eq:mapping00}
  C&=&\sqrt{1-\epsilon}\, a-\sqrt{\epsilon}\, b \, .
\end{eqnarray}
We choose such $\epsilon$ that the Hamiltonian~(\ref{eq:HamiltonianEFF}) can be
expressed in the form $H_{\rm{eff}}=-\delta_{r} C^{\dagger} C$, where 
$\delta_r=(4\delta_{2}^{2}+\delta_{1}^{2})^{1/2}$. Using this form of the Hamiltonian
one can derive the formula for the average photon number
\begin{eqnarray}
  \label{eq:mapping01}
  \langle b^{\dagger}b\rangle&=&n_{\rm{ph}}\,(1-\delta_{1}^{2}/\delta_{r}^{2})\sin^{2}(\delta_r t/2) \, .
\end{eqnarray}
In figure~\ref{fig:btb} we plot the average photon number as a function of time.
This figure shows that all  photons can be transferred from the first mode
to the second mode. However, this is possible only if $\delta_1=0$.
We want the state mapping to be
perfect, and therefore, we restrict ourselves to this case only. We can make $\delta_1\approx0$
by choosing values of $\Omega$ and $\Omega'$, which are much greater than $\Delta$ and satisfy condition $\Omega' g\approx\Omega g'$.
If one wants $\delta_1=0$ then values of $\Omega$ and $\Omega'$ have to be chosen more precisely
\begin{eqnarray}
  \label{eq:mapping02}
  \Omega'&=&\sqrt{(\Omega^2-2\Delta^2) g^{\prime 2}/g^2+2\Delta^2}\, .
\end{eqnarray}
For reference, we also calculate numerically the average photon number using the non-Hermitian Hamiltonian
\begin{eqnarray}
  \label{eq:HamiltonianNH}
  \tilde{H}&=&(\Delta-i\gamma'/2)\sigma_{11}+(\Delta-i\gamma/2)\sigma_{22}
  +(2\Delta-i\gamma''/2) \sigma_{33}+(\Omega\sigma_{23}+\Omega'\sigma_{13} +g a^{\dagger} \sigma_{02}
  +g' b^{\dagger} \sigma_{01}+ {\rm{h.c.}})\, ,
\end{eqnarray}
which governs the evolution of this open system during the time intervals when no collapse
occurs~\cite{pleniotrajek,carmichaelksiazka}. We have obtained the Hamiltonian~(\ref{eq:HamiltonianNH}) by substituting the relevant symbols in
\begin{eqnarray}
  \label{eq:nonHermitian}
  \tilde{H}&=&H-\frac{i}{2}\sum_{j} L_{j}^{\dagger}L_{j}\, 
\end{eqnarray}
with quantities from equations~(\ref{eq:Hamiltonian0}) and~(\ref{eq:L1}) for $n=1$.
As one can see from figure~\ref{fig:btb}, the analytical results are in a remarkable agreement with the numerical solution
even for quite considerable values of $\gamma$, $\gamma'$ and $\gamma''$ (where $\gamma''=\gamma_3+\gamma_3^{\prime}$)
as long as parameter regime justifies adiabatic elimination. 

\begin{figure}[ht]
\centering
  \includegraphics[width=6.5cm]{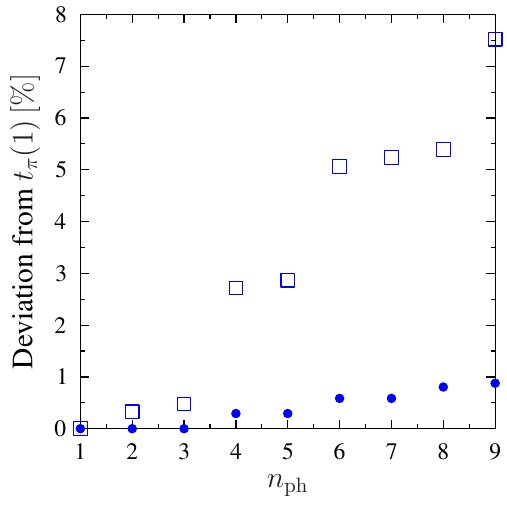}
   \caption{Deviation of $t_{\pi}(n_{\rm{ph}})$ from $t_{\pi}(1)$ (in percent) 
   for $(g',\Delta,\Omega,\Omega')/g=(1,10,33,33)$ (open squares) and for
   $(g',\Delta,\Omega,\Omega')/g=(1,30,100,100)$ (solid circles). The second parameter regime 
   justifies adiabatic elimination for $n_{\rm{ph}}=9$, whereas the first one
   only for $n_{\rm{ph}}=1$.}
  \label{fig:tpi_n}
\end{figure}

From equation~(\ref{eq:mapping01}) we can infer that the $\pi$ pulse time is given by the formula
\begin{eqnarray}
  \label{eq:mapping03}
t_{\pi}&=&\pi/\delta_{r}\, ,
\end{eqnarray}
from which one can observe one more important feature of the Hamiltonian.
It is evident that the time of such $\pi$ pulse is independent of $n_{\rm{ph}}$, and thus, we
are able to perform the state-mapping operation defined by 
$|n_{\rm{ph}}\rangle_{A}\otimes |0\rangle_{B}\to |0\rangle_{A}\otimes |n_{\rm{ph}}\rangle_{B}$.
Let us move into the rotating frame, in which the Hamiltonian takes the form
\begin{eqnarray}
  \label{eq:mapping04}
  H_{\rm{eff}}&=&-\delta_2 (a^{\dagger}a+b^{\dagger}b)
  +\delta_2 (a^{\dagger}b+b^{\dagger}a)\, 
\end{eqnarray}
and let us assume that the first mode is initially prepared in some interesting quantum
state $|\Psi_0\rangle=\sum_{k} c_{k}|k\rangle_{A}$, while the second mode is empty.
Then, by switching the lasers on for $t_{\pi}$, one can map this interesting state
onto the second mode
\begin{eqnarray}
  \label{eq:mapping05}
  \Big(\sum_{k} c_{k}|k\rangle_{A}\Big)\otimes |0\rangle_{B}&\to&
  |0\rangle_{A}\otimes \Big(\sum_{k} c_{k}|k\rangle_{B}\Big) \, .
\end{eqnarray}
In a frame rotating at different frequency, in which the Hamiltonian takes the form
\begin{eqnarray}
  \label{eq:mapping06}
  H_{\rm{eff}}&=&\delta_x (a^{\dagger}a+b^{\dagger}b)
  +\delta_2 (a^{\dagger}b+b^{\dagger}a)\, ,
\end{eqnarray}
phase factors appear and the $\pi$ pulse changes the initial state according to
\begin{eqnarray}
  \label{eq:mapping07}
  \Big(\sum_{k} c_{k}|k\rangle_{A}\Big)\otimes |0\rangle_{B}&\to&
  |0\rangle_{A}\otimes 
  \Big(\sum_{k} c_{k} e^{i\phi_{\pi}(k)}|k\rangle_{B}\Big) \, ,
\end{eqnarray}
where $\phi_{\pi}(n_{\rm{ph}})=-n_{\rm{ph}}\pi(\delta_2+\delta_x)/(2\delta_2)$.
Note that for the parameters values used in figure~\ref{fig:btb}
$\delta_{0}=-2.74$ and $\delta_{2}=2.98$, so $\delta_{0}\approx -\delta_{2}$.
For the Hamiltonian~(\ref{eq:HamiltonianEFF}), $\delta_{1}=0$ and large $\Omega$
there are no phase factors, because $\delta_{0}$ tends to $-\delta_{2}$ for large $\Omega$,
and thus, the Hamiltonian~(\ref{eq:HamiltonianEFF}) tends to the form given by equation~(\ref{eq:mapping04}).
The independence of $t_{\pi}$ from $n_{\rm{ph}}$ is crucial for the state-mapping operation.
Unfortunately, $t_{\pi}$ is independent of $n_{\rm{ph}}$ only in the approximated
model~(\ref{eq:HamiltonianEFF}), in which we adiabatically eliminated all atomic excited levels.
Numerical calculations show that $t_{\pi}$ increases with $n_{\rm{ph}}$
in the more general model of the system given by the Hamiltonian~(\ref{eq:Hamiltonian0}) for $n=1$.
However, as long as the adiabatic elimination is justified, we can neglect the dependence
$t_{\pi}$ on $n_{\rm{ph}}$, as is seen in figure~\ref{fig:tpi_n}. 
It is seen from figure~\ref{fig:tpi_n} that there are jumps of the value of $t_{\pi}$.
These jumps come from the fact that populations of atomic excited levels oscillate
with high frequencies~\cite{chimczak07,chimczak08,chimczak15}. Thus, there are many local
closely-spaced maxima of the population of the desired final state $|0\rangle_{A}\otimes |n_{\rm{ph}}\rangle_{B}$.
Therefore, the global maximum ($t_{\pi}$) changes sometimes discontinuously with increasing
of $n_{\rm{ph}}$ --- from one local minimum to the next one. We can neglect these jumps as
long as the adiabatic elimination is justified.

Let us now investigate the effect of $\gamma$ and $\gamma'$ on the state-mapping operation.
To this end we need non-Hermitian Hamiltonian, which we obtain by inserting Eq.~(\ref{eq:HamiltonianEFF}) and the relevant effective rotating frame Lindblad operators (see Reiter-S{\o}rensen method in Methods)  into equation~(\ref{eq:nonHermitian}). Assuming that $\Omega,\Omega'\gg\Delta$ and $\delta_{1}=0$, this Hamiltonian can be quite well approximated by
\begin{eqnarray}
  \label{eq:nHCtC}
  \tilde{H}=-2\delta_{2} C^{\dagger} C-\frac{i}{2}\gamma_{\rm{eff}} C^{\dagger} C\, ,
\end{eqnarray}
where the effective dissipation rate is given by
\begin{eqnarray}
  \label{eq:nHCtC2}
   \gamma_{\rm{eff}}&=&\frac{2 n g^2 g^{\prime 2}(g^2\gamma' +g^{\prime 2}\gamma)}
   {\Delta^2(g^2+g^{\prime 2})^2}\, .
\end{eqnarray}
It is clear that the fidelity of the state mapping ${\cal{F}}$ and the probability that
no collapse occurs during this operation ${\cal{P}}$ are close to one only if
the effective dissipation rate $\gamma_{\rm{eff}}$ is much less than the effective coupling
strength $\delta_{2}$.
For $g=g'$ and $\gamma=\gamma'$, the expression for the effective dissipation
rate takes the simpler form $\gamma_{\rm{tot}}=n\gamma g^2/\Delta^2$. In this special
case, $\cal{F}$ and $\cal{P}$ depend on the ratio $\gamma/\Delta$.
Let us now check this result numerically using the non-Hermitian Hamiltonian
\begin{eqnarray}
  \label{eq:HamiltonianNHsum}
  \tilde{H}=\sum_{k=1}^{n}\Big\{(\Delta-i\gamma'/2)\sigma_{11}^{(k)}
  +(\Delta-i\gamma/2)\sigma_{22}^{(k)}+(2\Delta-i\gamma''/2) \sigma_{33}^{(k)}
  +(\Omega\sigma_{23}^{(k)}+\Omega'\sigma_{13}^{(k)}+g a^{\dagger} \sigma_{02}^{(k)}
  +g' b^{\dagger} \sigma_{01}^{(k)}+ {\rm{h.c.}})\Big\}\, .
\end{eqnarray}
First, we have to choose specific values of parameters. The choice of the atom-cavity system
determines $g$, $g'$, $\gamma$, $\gamma'$ and $\gamma''$. For macroscopic cavities
$g/2\pi$ is typically of the order of $10$~MHz and $\gamma$ ranges from about $0.2 g$ to
$g$~\cite{boozerPRL07,wilkSCI07}. Let us set $g'=g=2\pi\cdot 10$~MHz, 
$\gamma'=\gamma=2 g$ and $\gamma''=g$. The choice of the initial state determines the 
Fock state $|n_{\rm{ph}}\rangle$, to which the state mapping has to be faithful.
Let the initial state of the $a$ mode be $|\Psi_0\rangle=(|0\rangle_{A}+|1\rangle_{A}+|2\rangle_{A}+|3\rangle_{A})/2$.
If there are four atoms trapped in the cavity, then the detuning has to satisfy 
$\Delta\gg g \sqrt{4\cdot 3}$. We set $\Delta=35 g$.
Finally, we choose the value of $\Omega$ and calculate $\Omega'$ using equation~(\ref{eq:mapping02}).
These values have to be large enough to satisfy the second condition in~(\ref{eq:condA}).
It is easy to check that for $\Omega=\Omega'=175 g$ this condition is fulfilled, and therefore,
adiabatic elimination is justified.
For $(g',\Delta,\Omega,\Omega',\gamma,\gamma',\gamma'')/g=(1,35,175,175,2,2,1)$ and $n=4$ we
have found that ${\cal{F}}=0.993$ and ${\cal{P}}=0.885$. In the case of one atom trapped in
the cavity ($n=1$), for the same parameters, we have found that ${\cal{F}}=0.995$ and ${\cal{P}}=0.886$.
One can see that ${\cal{F}}$ and ${\cal{P}}$ are almost the same in the two cases. The only
important difference is the time of the state-mapping operation --- $t_{\pi}=26.5/g$ and $t_{\pi}=105.6/g$
for $n=4$ and $n=1$, respectively. The time of the state mapping in the one-atom case
is almost four times larger than that in the four-atom case. This result is in an agreement
with equation~(\ref{eq:mapping03}). We can make $t_{\pi}$ smaller in the one-atom case by setting smaller
$\Delta$ but then the ratio $\gamma/\Delta$ increases and the dissipation reduces the fidelity and the
success probability. For instance, if we set $(g',\Delta,\Omega,\Omega',\gamma,\gamma',\gamma'')/g=(1,17,85,85,2,2,1)$
in the one-atom case then the time of the state mapping is reduced to $t_{\pi}=51.6/g$. Then, however
the dissipation reduces the fidelity and the success probability to ${\cal{F}}=0.979$ and ${\cal{P}}=0.795$, 
respectively. 
%------------------------------------------------------------------

%\section*{Fast opening high-{\cal{Q}} cavity system}\label{sec:device}

\vspace*{5mm}
\noindent
\textbf{Quantum-state extraction by fast opening high-{\cal{Q}} cavity.}
The investigated system can be applied as
a fast opening high-{\cal{Q}}
cavity  that can be easily and coherently controlled with classical laser fields.
The device is based on similar principles as the setup of Tufarelli~\emph{et~al.}~\cite{tufarelliPRL14},
but it employs four-level atoms in the diamond configuration instead of two-level atoms.
The main idea of both setups is to couple a high-{\cal{Q}} cavity mode to a low-{\cal{Q}} cavity mode
through atoms. Such a device would be very useful, because on the one
hand we need a high {\cal{Q}} factor to reach the strong coupling regime~\cite{hoodSCI00,pinkseNAT00,
mckeeverNAT03,boozerPRL07,reisererSCI13,reisererNAT14,reisererRMP15,nollekePRL13},
in which we can generate a complex non-classical state of light trapped inside optical 
resonator~\cite{larsonJOM06,pradoPRA11,domokosEPJD98,savage90,rongChPB11,nikoghosyanPRL12,
liuQIP13,ficekPRA13,xiaoQIP13,liuQIP14}. On the other hand, we need a low {\cal{Q}} factor
to extract this state from the resonator into a waveguide before it will be distorted by
the cavity damping. The device proposed by Tufarelli~\emph{et~al.}~\cite{tufarelliPRL14}
makes it possible to change the effective {\cal{Q}} factor. If atoms are absent, there is no coupling
between the two modes and the whole system works as an effective high-{\cal{Q}} cavity. If we move atoms
into the cavity, then photons leak out of the high-{\cal{Q}} mode through the low-{\cal{Q}} mode and the
whole device works as an effective low-{\cal{Q}} cavity. Instead of shifting the atoms out of the cavity
we can shift atoms out of resonance using a laser and the dynamic Stark effect.
As long as the laser illuminates atoms, there is no coupling between modes.
Here, we propose to replace two-level atoms by four-level atoms in the diamond configuration.
Our modification allows us to use a bimodal cavity, 
which supports circularly polarised modes of the same or different polarisations and frequencies. 
Moreover, it requires intense laser light to illuminate atoms only in short
time intervals, when we need the coupling between modes. When the laser is switched off, there
is no coupling between modes.

%=============================================
\section*{Discussion}
\begin{figure}[ht]
\centering
  \includegraphics[width=8cm]{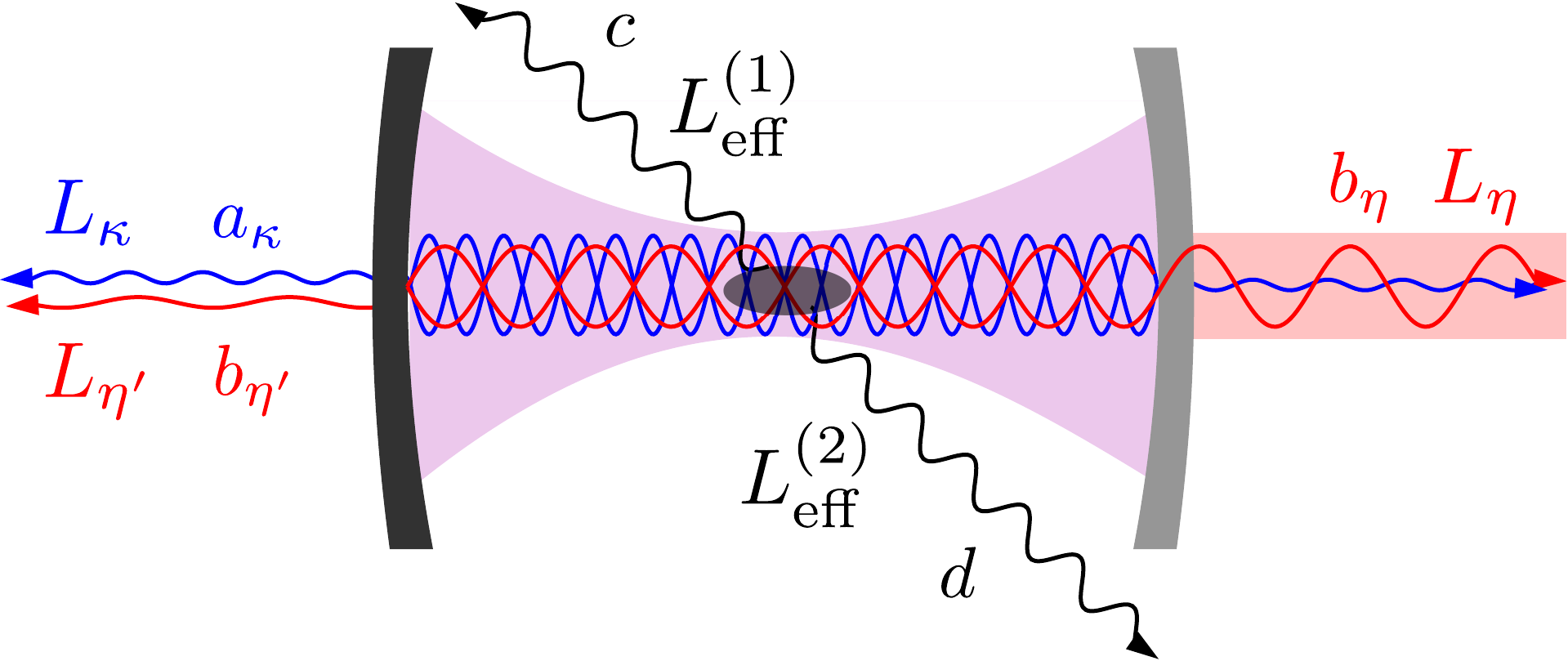}
  \caption{The effective model of the setup and the waveguide. The photon emission
  from the open cavity into the waveguide is represented by the operator $L_{\eta}$. 
  All photon losses are modelled by four inaccessible travelling modes. Emissions
  to these modes are described by $L_{\eta'}$, $L_{\kappa}$, 
  $L_{\rm{eff}}^{(1)}$ and $L_{\rm{eff}}^{(2)}$.}
  \label{fig:systemEFF}
\end{figure}

%\subsection{Performance analysis}
After the adiabatic elimination of atomic excited states we can restrict our
considerations to a simplified model, which does not include atomic variables. Such
simplified model makes it easy to take into account all photon losses. To this
end, we model the device as two cavity modes, which decay emitting the radiation
into five travelling modes, as is depicted in figure~\ref{fig:systemEFF}.
One of these travelling modes is accessible experimentally. This accessible travelling
mode can be, for example, a waveguide. Other travelling modes are inaccessible,
and thus, provide losses.
The photon emissions from both cavity modes (represented by operators $a$ and $b$)
into the inaccessible travelling modes are described by the Lindblad operators:
$L_{\eta'}=\sqrt{\eta'}b$, $L_{\kappa}=\sqrt{\kappa}a$, $L_{\rm{eff}}^{(1)}$ and
$L_{\rm{eff}}^{(2)}$. The photon emission into the accessible travelling mode is 
described by the Lindblad operator $L_{\eta}=\sqrt{\eta}b$.
Here we assume, unless explicitly stated otherwise, that the device is working
in {\emph{the open mode}}, \emph{i.e.}, both lasers are turned on ($\Omega,\,\Omega'\neq 0$). 
We also assume that the quantum state of field was prepared in advance in 
the mode represented by the operator $a$.
Under these assumptions, we derived a quantity that describes
the quality of the field extracted from the resonator into a waveguide. 
We refer to this quantity as to the figure of merit of the proposed device (see Methods). Let us now investigate the usefulness of the considered device to extract
a field state from the $a$ mode.
We assume that there is only one optical cavity. This cavity supports
two electromagnetic field modes of different frequencies $\omega$ and $\omega'$
(see figure~\ref{fig:systemEFF}). The first of them is considered as the
$a$ mode, while the second one as the $b$ mode.
Each cavity mirror is described by its radius of
curvature $r$, transmission coefficients $T$ and $T'$ for the $a$ mode and the 
$b$ mode, respectively, and loss coefficient $L$, which is assumed to be the
same for both modes.
The $a$ mode requires very low values of $T$ and $L$ for both mirrors.
To our knowledge, these parameters take the lowest value for the mirror
that has been used in the experiment of Refs.~[\citenum{wilkSCI07,wilk2008Phd}].
We set these values in our calculations, \emph{i.e.}, $T_{\rm{small}}=T_1=T_2=T'_1=1.8$~ppm
and $L=3.15$~ppm, where the subscripts indicate the mirror.
The radius of curvature of both mirrors is 50~mm~\cite{wilkSCI07,wilk2008Phd}.
Now we can vary only the cavity length $l$ and the transmission coefficient
$T'_2$, and therefore, we want to plot the figure of merit $F$ as a 
function of these two quantities. First, we have to choose a concrete
realisation of the $\Diamond$-type atom.
Let us choose a ${}^{87}{\rm{Rb}}$ atom, as in the mentioned above
experiment~\cite{wilkSCI07,wilk2008Phd}, and
its levels $|5 S_{1/2}, F=2, m_{F}=2\rangle$, $|5 P_{3/2}, F=3, m_{F}=3\rangle$,
$|6 P_{3/2}, F=3, m_{F}=3\rangle$ and $|6 D_{3/2}, F=3, m_{F}=3\rangle$ to
serve as $|0\rangle$, $|1\rangle$, $|2\rangle$ and $|3\rangle$, respectively.
This choice determines values of modes frequencies to be $\omega/2\pi=713.28$~THz 
and $\omega'/2\pi=384.23$~THz~\cite{kramida2015nist}. The lifetimes of
all used here excited levels can be found in Ref.~[\citenum{safronovaPRA11}].
It is important that the lifetime of the level $|3\rangle$ is longer
($\tau_3=256$~ns) than lifetimes of the other excited levels
($\tau_1=112$~ns for $|1\rangle$ and $\tau_2=26.25$~ns for $|2\rangle$).
So our assumption that spontaneous emissions from the excited level
$|3\rangle$ can be neglected in calculations is justified not only
by small population of this level, but also by $\tau_3>\tau_1$ and
$\tau_3\gg\tau_2$. The spontaneous emissions can take the $^{87}\mathrm{Rb}$ atom
from the states $|1\rangle$ and $|2\rangle$ only to state $|0\rangle$.
Hence, it is easy to calculate corresponding spontaneous emission
rates: $\gamma/2\pi=1.42$~MHz and $\gamma'/2\pi=6.06$~MHz.
In principle, the scheme presented here works properly even with only
one trapped atom. In real experiments, however, this scheme will require
a much larger number of atoms to achieve the figure of merit that is
close to unity. In order to compare our scheme with the original scheme of
Tufarelli~\emph{et al.}~\cite{tufarelliPRL14}, we set here the same number
of atoms as in Ref.~[\citenum{tufarelliPRL14}], \emph{i.e.}, $n=1000$. Trapping $1000$ rubidium
atoms and preparing them in the $|5 S_{1/2}, F=2, m_{F}=2\rangle$ state 
is possible using fiber-based Fabry-Perot cavities~\cite{colombeNat2007, Haas1SCI14}.
We have chosen the macroscopic cavity in our considerations. A number of atoms
trapped inside macroscopic cavities is typically of the order of
$10^{5}$~\cite{BrenneckeSci08}. Trapping ${\sim}1000$ atoms also should be
possible.
Now we can calculate the coupling strength $g$ using
\begin{eqnarray}
  \label{eq:g} 
  g&=&\sqrt{\frac{3\pi c^3\gamma}{2\omega^2 V}}\, ,
\end{eqnarray} 
where $c$ is the speed of light and $V$ is the cavity mode volume given by
\begin{eqnarray}
  \label{eq:V} 
  V&=&\pi c l\sqrt{l(2 r-l)}/(4\omega)\, .
\end{eqnarray}
In order to calculate the coupling strength $g'$ we have to replace
$\omega$ and $\gamma$ by $\omega'$ and $\gamma'$ in~(\ref{eq:g})
and~(\ref{eq:V}).
The cavity damping constants of the considered scheme can be calculated as
\begin{eqnarray}
  %\label{eq:Kappa} 
  \kappa=c(1-R)/(l\sqrt{R})\, ,\quad
  %\label{eq:Eta} 
  \eta=T'_2\eta_{\rm{tot}}/{\cal{N}}\, ,\quad
  \label{eq:Etaprime} 
  \eta'=(2 L+T'_1)\eta_{\rm{tot}}/{\cal{N}}\, ,
\end{eqnarray} 
where $R=1-L-T_{\rm{small}}$, ${\cal{N}}=2 L+T'_1+T'_2$ and
\begin{eqnarray}
  \label{eq:Etatot} 
  \eta_{\rm{tot}}&=&c\frac{1-\sqrt{R(1-L-T'_2)}}{l R^{1/4}(1-L-T'_2)^{1/4}}\, .
\end{eqnarray}  
Finally, we have to fix values of $\Delta$, $\Omega$ and $\Omega'$.
It is necessary to choose these values carefully. On the one hand,
they should be big enough to make adiabatic elimination justified.
On the other hand, they cannot be too big, because $\delta_2$ and extraction efficiency
decrease with increasing $\Delta$. In our computations we set
$\Delta/g=700$ and $\Omega/\Delta=5$, which justifies adiabatic
elimination for cavity states with $\langle a^{\dagger} a\rangle\lesssim 10$.
Then $\Omega'$ is given by~(\ref{eq:mapping02}).

\begin{figure}[ht]
\centering
  \includegraphics[width=8cm]{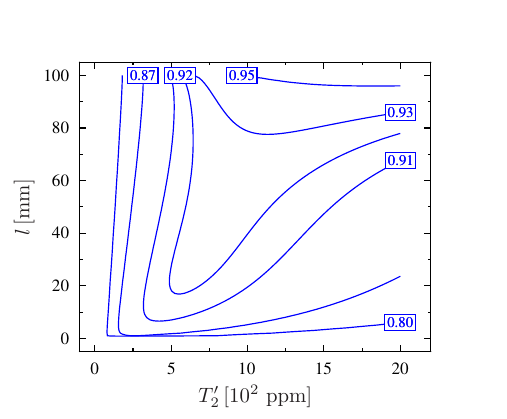}
  \caption{Figure of merit $F$ as a function of the cavity length $l$ and
           the transmission coefficient $T'_2$ for $1000$ $\Diamond$-type 
           four-level atoms.}
  \label{fig:resultsF}
\end{figure}

\begin{figure}[ht]
\centering
  \includegraphics[width=8cm]{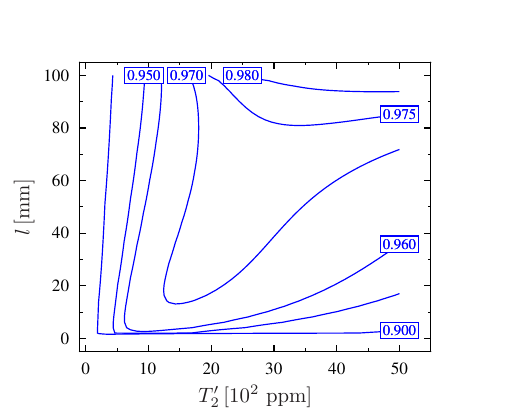}
  \caption{Figure of merit $F$ as a function of the cavity length $l$ and
           the transmission coefficient $T'_2$ for $8000$ $\Diamond$-type 
           four-level atoms.}
  \label{fig:F_n}
\end{figure}

%\textbf{Experimental considerations}
Now, we can discuss the experimental feasibility of the scheme. In order to do so let us use figure of merit
$F$ closely related to the probability of successful
operation of the discussed device.
Under certain conditions (given explicitly in Methods) $F$ is given as
\begin{eqnarray}
  \label{eq:Ffinal} 
  F=\frac{\eta}{\eta_{\rm{tot}}}\Big[1
-\frac{(\sqrt{\zeta_1\theta_1}+\sqrt{\zeta_2\theta_2})^2}{2\delta_2^2}-\frac{\eta_{\rm{tot}}\big(\kappa+\zeta_1+\zeta_2\big)}
{4\delta_2^2+\eta_{\rm{tot}}\big(\kappa+\zeta_1+\zeta_2\big)}\Big]\, ,
\end{eqnarray} 
where  $\zeta_1=n\gamma'\alpha_3^2 g^2$, $\theta_1=n\gamma'\alpha_1^2 g^{\prime 2}$,
$\zeta_2=n\gamma\alpha_2^2 g^2$, $\theta_2=n\gamma\alpha_3^2 g^{\prime 2}$,
$\eta_{\rm{tot}}=\eta'+\eta$.
We can  plot $F$ as a function of $l$ and $T'_2$.
For the parameters given above the formula~(\ref{eq:Ffinal}) gives only raw
approximation of the figure of merit. 
Therefore, we have calculated $F$
numerically using its definition (given in Methods), and we have obtained in this
way results presented in figure~\ref{fig:resultsF}.
As expected, the figure of merit takes the maximum value in the near-concentric
regime $l\approx 2\,r$. For $l=99.9$~mm and $T'_2=2000$~ppm the figure of merit
is equal to $0.97$. Unfortunately, the near-concentric configuration of
the macroscopic mirrors is extremely sensitive to misalignment, and therefore,
it would be difficult or even impossible to achieve such high value of
$F$~\cite{hungerNJP10}. For the confocal configuration $l=r$, which is the most
stable configuration, the figure of merit can be equal to $0.92$. 
This value is still quite high and it is higher than $F$ of
the original scheme of Tufarelli~\emph{et al.}~\cite{tufarelliPRL14}.
Of course, we can always increase the figure of merit by increasing $n$.
To show this we plot also the figure of merit for $n=8000$.
It is seen from figure~\ref{fig:F_n} that now $F=0.97$ even for the confocal
configuration.
Form equation~(\ref{eq:Ffinal}) it follows that the figure of merit can be close to one only under the condition
$4\delta_2^2\gg\eta_{\rm{tot}}\big(\kappa+\zeta_1+\zeta_2\big)$.
Assuming $\delta_1=0$, this condition can be expressed as:
\begin{eqnarray}
  \label{eq:conds} 
  \eta_{\rm{tot}}\ll n g^{2}/\gamma\, , & \eta_{\rm{tot}}\ll n g^{\prime 2}/\gamma'\, , &
   \eta_{\rm{tot}}\ll\delta_2(\delta_2/\kappa)\, .
\end{eqnarray} 
It follows that $\kappa$ has to be at least two orders of magnitude smaller
than $\delta_2$. For currently available atom-cavity systems all these conditions
can be satisfied only for a large number of atoms $n$.
Note that $F$ is independent of $\delta_1$ in the mentioned regime. Typically,
$g\neq g'$ in concrete realisations of the four-level atom, and therefore,
usually $\delta_1\neq 0$. A non-zero value of $\delta_1$ decreases $F$ when
dissipative rates are too large. It is possible to make $\delta_1=0$ by
setting appropriate value of $\Omega'$, \emph{i.e.}, this one given by 
equation~(\ref{eq:mapping02}).
From equation~(\ref{eq:Ffinal}), it is seen that such precise setting of $\Omega'$
is not necessary in the regime, in which the figure of merit is close to one.
This feature makes choosing values of parameters easier. It is also worth to
note that in the regime $\Omega\gg\Delta$ equation~(\ref{eq:mapping02}) takes the simpler
form $\Omega' g\approx\Omega g'$.
Let us now verify the approximate formula~(\ref{eq:Ffinal}) for the parameter regime corresponding to the confocal configuration with $1000$ atoms. By setting $l=50$~mm and $T'_2=800$~ppm, we get $(g, g', \Delta, \Omega, \Omega', \gamma, \gamma', \gamma'',\eta_{\rm{tot}},\kappa)/(2\pi)=(0.1, 0.29, 72.8, 364, 989, 1.4, 6.06, 0.6, 0.4, 4.7\cdot 10^{-3})$~MHz. These lead to $(\zeta_1, \theta_1, \zeta_2, \theta_2)/(2\pi)=$
$(1.3, 2.3, 1.2, 2.5)$~kHz and $\delta_2/(2\pi)=0.14$~MHz. As mentioned earlier,
the formula~(\ref{eq:Ffinal}) is valid if the conditions
$\delta_2\gg\kappa\, , \zeta_1\, , \theta_1\, , \zeta_2\, , \theta_2$ and 
$\eta_{\rm{tot}}\gg\delta_2$ are fulfilled. One can see that the first condition
is fulfilled. However, the ratio $\eta_{\rm{tot}}/\delta_2$ is only $2.8$. Nevertheless,
the value of the figure of merit calculated using equation~(\ref{eq:Ffinal}), i.e., $F=0.95$ is
quite close to the value $F=0.92$ obtained numerically using its definition.

%=============================================

%\subsection{The device working in the closed mode}
So far, we have investigated the device working in the open mode. Let us
now consider this device working in the closed mode. 
For the device working in the closed mode both lasers
are turned off. The effective Hamiltonian derived with $\Omega=\Omega'=0$
is given by equation~(\ref{eq:Hamiltonian_closed}).
It is seen that there is no interaction between the $a$ mode and the $b$ mode,
and therefore, photons do not leak out of the $a$ mode through the $b$ mode.
The only destructive role played by atoms trapped inside the cavity is the increase
of photon losses caused by the spontaneous emission from the atomic excited
state $|2\rangle$. 
The decay of the $a$ mode associated with the atomic spontaneous emission is 
described by an effective decay rate $\kappa_{\gamma}\approx n \gamma(g/\Delta)^2$
[see equation~(\ref{eq:EFF0})].
We have found out that for the parameters values used above
($l=50$~mm, $T'_2=800$~ppm and $n=1000$) this effective decay rate 
$\kappa_{\gamma}/(2\pi)=2.9\cdot 10^{-3}$~MHz is less than the cavity
decay rate associated with the absorption in the mirrors
$\kappa/(2\pi)=4.7\cdot 10^{-3}$~MHz. Knowing $\kappa_{\gamma}$, we can take
atomic spontaneous emissions into account just by making the replacement
$\kappa\to\kappa'=\kappa+\kappa_{\gamma}$.

%------------------------------------------------------------------

\section*{Conclusion}

We have studied a quantum system composed of $\Diamond$-type atoms
and an optical cavity supporting two electromagnetic field modes, in which
these atoms are permanently trapped. We have considered the case, where
lower atomic transitions (see figure~\ref{fig:diamond}) are coupled to the field modes and 
upper atomic transitions are driven by classical laser fields. We have shown that this 
complex quantum system can be described by an effective Hamiltonian of the simple
form given in equation~(\ref{eq:HamiltonianEFF}) if intensities of the lasers
fields and the detuning are sufficiently large. We have also shown that the evolution
of this system can be easily controlled just by turning lasers on and off.
We have presented two examples of applications of the system.
The first application is a state transfer from one quantized mode to another.
We have shown that the time of the state transfer is independent of the number of photons.
Thus, it is possible to map a quantum state of one mode onto the other mode.
As the second application of the system, we have presented a device which can be switched on demand to perform either as a low-{\cal{Q}} cavity, or as a high-{\cal{Q}} cavity. The $\Diamond$-type atoms allow
for fast switching between these two working modes just by switching the lasers on
and off. Moreover, $\Diamond$-type atoms make this device to be especially
well suited for a bimodal cavity, which supports circularly polarised modes
of the same or different polarisations and frequencies.

\section*{Methods}

%\subsection{Adiabatic elimination of excited atomic states}
\noindent
\textbf{Reiter-S{\o}rensen method.} Initially, all atoms are prepared in the ground state. An atom can be found
in one of the excited states, only if it absorbs a single photon. 
We want to achieve an effective coupling between field modes and
no coupling between the modes and atoms. Therefore, the atomic excited 
states have to be populated only virtually. In this case, we can adiabatically
eliminate the atomic excited states and use in calculations an effective Hamiltonian 
for the ground state subspace. To this end, we use the effective operator formalism 
for open quantum systems described in Ref.~[\citenum{reiterPRA12}]. 
Let us consider the single atom case first.
The Hamiltonian describing a single atom can be easily obtained by simplifying
equation~(\ref{eq:Hamiltonian0}) and it reads
\begin{equation}
  \label{eq:Hamiltonian1}
  H=\Delta \sigma_{11}+\Delta \sigma_{22}+
  2\Delta \sigma_{33}+(\Omega\sigma_{23}+
  \Omega'\sigma_{13} +g a^{\dagger} \sigma_{02}+g' b^{\dagger} \sigma_{01}
  + {\rm{h.c.}})\, .
\end{equation}
The Lindblad operators representing spontaneous transitions from the atomic
excited states are given by
\begin{eqnarray}
  \label{eq:L1b}
  L_{1}=\sqrt{\gamma'}\sigma_{01}\, , \quad
  L_{2}=\sqrt{\gamma}\sigma_{02}\, , \quad
  L_{3}=\sqrt{\gamma_{3}}\sigma_{23}\, ,\quad
  L_{4}=\sqrt{\gamma'_{3}}\sigma_{13}\, ,  
\end{eqnarray}
where $\gamma$, $\gamma'$, $\gamma_{3}$ and $\gamma'_{3}$ are spontaneous
emission rates for the respective transitions. The master equation of
Kossakowski-Lindblad form describing the evolution of this system
is then given by
\begin{equation}
  \label{eq:master}
  \dot{\rho}=-i [H,\rho]  +\sum_{j=1}^{4} \Big[L_{j}\rho L_{j}^{\dagger}
  -\frac{1}{2} (L_{j}^{\dagger}L_{j}\rho  +\rho L_{j}^{\dagger}L_{j})\Big]\, .
\end{equation}
The effective-operator formalism for open quantum systems~\cite{reiterPRA12}
reduces equation~(\ref{eq:master}) to an effective master equation, where the dynamics
is restricted to the atomic ground state only. 
In order to apply the effective-operator formalism, we need to 
provide: the Lindblad operators, the Hamiltonian in the exited-state 
manifold $H_{\rm{e}}$, the ground-state Hamiltonian $H_{\rm{g}}$ (here $H_{\rm{g}}=0.$), and the perturbative \mbox{(de-)excitations}
of the system $V_{+}$ ($V_{-}$). 
These are given by
\begin{eqnarray}
  \label{eq:eoffoqs}
  H_{\rm{e}}=\Delta \sigma_{11}+\Delta \sigma_{22}+
  2\Delta \sigma_{33}+(\Omega\sigma_{23}+\Omega'\sigma_{13} + {\rm{h.c.}})\, ,\quad 
  V_{-}=g a^{\dagger} \sigma_{02}+g' b^{\dagger} \sigma_{01}\, ,\quad
  V_{+}=g a \sigma_{20}+g' b \sigma_{10}\,.
\end{eqnarray}
The effective Hamiltonian and collapse operators can be derived
using formulas~\cite{reiterPRA12}
\begin{eqnarray}
  \label{eq:eoffoqs2}
  H_{\rm{eff}}&=&-\frac{1}{2}V_{-}[H_{\rm{NH}}^{-1}+(H_{\rm{NH}}^{-1})^{\dagger}]
  V_{+}+H_{\rm{g}}\, ,\\
  \label{eq:eoffoqs2b}
  L_{\rm{eff}}^{(j)}&=&L_{j} H_{\rm{NH}}^{-1} V_{+}\, ,
\end{eqnarray}
where
\begin{eqnarray}
  \label{eq:eoffoqs3}
  H_{\rm{NH}}&=&H_{\rm{e}}-\frac{i}{2}\sum_{j} L_{j}^{\dagger}L_{j}\, .
\end{eqnarray}
Assuming that all spontaneous emission rates are negligibly small compared with 
$\Omega$, $\Omega'$ and $\Delta$ we can approximate $H_{\rm{NH}}^{-1}$ by
\begin{eqnarray}
  \label{eq:eoffoqs4}
  H_{\rm{NH}}^{-1}\approx\alpha_1\sigma_{11}+\alpha_2\sigma_{22}
  +\alpha_3(\sigma_{12}+\sigma_{21})+\alpha_4\sigma_{33}
  +\alpha_5(\sigma_{13}+\sigma_{31})+\alpha_6(\sigma_{23}+\sigma_{32})\, ,
\end{eqnarray}
where $\alpha_1=\xi (\Omega^2-2\Delta^2)$,
$\alpha_2=\xi (\Omega^{\prime 2}-2\Delta^2)$,
$\alpha_3=-\xi\Omega\Omega'$, $\alpha_4=-\xi\Delta^2$, $\alpha_5=\xi\Delta\Omega'$,
$\alpha_6=\xi\Delta\Omega$ with 
$\xi=1/(\Delta[\Omega^2+\Omega^{\prime 2}-2\Delta^2])$.
From the combined equation~(\ref{eq:eoffoqs4}) and equation~(\ref{eq:eoffoqs2}) we derive
the effective Hamiltonian
\begin{eqnarray}
  \label{eq:HamiltonianEFFb}
  H_{\rm{eff}}&=&\delta_0 (a^{\dagger}a+b^{\dagger}b)+
  \delta_1 b^{\dagger}b+\delta_2 (a^{\dagger}b+b^{\dagger}a)\, ,
\end{eqnarray}
where $\delta_0=-g^2 \alpha_2$, $\delta_1=g^2\alpha_2-g^{\prime 2}\alpha_1$
and $\delta_2=- g g' \alpha_3$.
By inserting equation~(\ref{eq:eoffoqs4}) into equation~(\ref{eq:eoffoqs2b}) 
we obtain the effective Lindblad operators
\begin{eqnarray}
  \label{eq:Leff1}
  L_{\rm{eff}}^{(1)}=\sqrt{\gamma'}
  [\alpha_3 g a+\alpha_1 g' b]\, ,\quad
  L_{\rm{eff}}^{(2)}=\sqrt{\gamma}
  [\alpha_2 g a+\alpha_3 g' b]\, .
\end{eqnarray}
Unfortunately, deriving the expressions for operators $L_{\rm{eff}}^{(3)}$
and $L_{\rm{eff}}^{(4)}$ is more challenging than deriving $L_{\rm{eff}}^{(1)}$
and $L_{\rm{eff}}^{(2)}$.
First of all, the action of the operators $L_{3}$
and $L_{4}$ takes the system state to one of the excited states $|1\rangle$ or
$|2\rangle$, while all the excited states should be populated only virtually.
In the single atom case after spontaneous emission from the excited state $|3\rangle$
it is necessary
to reset the device, otherwise it will not work properly. Second, the effective 
operator formalism assumes that the excited states decay to the ground states only.
We circumvent these obstacles by choosing such values of parameters that
probabilities of occurrence of collapses described by $L_{3}$ and $L_{4}$ are
negligibly small. We will give later conditions for the parameters, which
allow us to neglect $L_{3}$ and $L_{4}$.
Using this approximation, we can write the effective master equation as
\begin{eqnarray}
  \label{eq:masterEFFb}
  \dot{\rho}=-i [H_{\rm{eff}},\rho]
  +\sum_{j=1}^{2} \left\lbrace L_{\rm{eff}}^{(j)}\rho (L_{\rm{eff}}^{(j)})^{\dagger}
  -\frac{1}{2} \left[L_{\rm{eff}}^{(j)})^{\dagger}L_{\rm{eff}}^{(j)}\rho
  +\rho (L_{\rm{eff}}^{(j)})^{\dagger}L_{\rm{eff}}^{(j)}\right]\right\rbrace\, .
\end{eqnarray}
It is easy to generalise this result to the $n$ atom case. In this more general
case the Hamiltonian is still given by equation~(\ref{eq:HamiltonianEFFb}), but with
\begin{eqnarray}
  \label{eq:d0d1d2n}
\delta_0=-n g^2 \alpha_2\, ,\,  \delta_2=-n g g' \alpha_3\, ,\quad {\mbox{and}} \quad
\delta_1=n (g^2\alpha_2-g^{\prime 2}\alpha_1)\, .
\end{eqnarray}
For the sake of simplicity, we have assumed here
that coupling strengths $g$ and $g'$ are the same for each atom in the ensemble. 
Note, however, that every atom in a Bose-Einstein condensate indeed experiences an
identical coupling to the cavity mode~\cite{tufarelliPRL14}.

In a frame rotating at $\delta_0$ the Hamiltonian takes the form
\begin{eqnarray}
  \label{eq:HamiltonianEFF2}
  H_{\rm{eff}}&=&\delta_1 b^{\dagger}b+\delta_2 (a^{\dagger}b+b^{\dagger}a)\, .
\end{eqnarray}
The effective Lindblad operators are now given by
\begin{eqnarray}
  \label{eq:Leff2}
  L_{\rm{eff}}^{(1)}=\sqrt{n \gamma'}
  [\alpha_3 g a+\alpha_1 g' b]\, \quad
  L_{\rm{eff}}^{(2)}=\sqrt{n \gamma}
  [\alpha_2 g a+\alpha_3 g' b]\, .
\end{eqnarray}

\vspace{5mm}
\noindent
\textbf{Alexanian-Bose method.} Using the effective operator formalism and assuming
that the upper level $|3\rangle$ can be neglected, 
we have obtained all needed formulas. Unfortunately we still do not
know the limits in which these approximations are valid.
In order to determine the limits we derive the effective Hamiltonian~(\ref{eq:HamiltonianEFF})
using another method --- a perturbative unitary transformation~\cite{alexanianPRA95}.
An incidental bonus is that this method provides new insights into the dynamics of
the four-level atom in the diamond configuration. Let us start by decomposing the 
Hamiltonian~(\ref{eq:Hamiltonian1}) into two parts
\begin{eqnarray}
  \label{eq:HamiltonianA1}
  H&=&H_0+H_1\, ,
\end{eqnarray}
where
\begin{eqnarray}
  \label{eq:HamiltonianA1B}
  H_0&=&\Delta \sigma_{11}+\Delta \sigma_{22}+
  2\Delta \sigma_{33}+(\Omega\sigma_{23}
  +\Omega'\sigma_{13}+ {\rm{h.c.}})\, .
\end{eqnarray}
Diagonalizing $H_0$ in the basis $\{ |1\rangle, |2\rangle, |3\rangle\}$
leads to the dressed states energies
\begin{eqnarray}
  \label{eq:HamiltonianA2}
  \Delta\, , &\quad (3\Delta-\Omega_{\rm{R}})/2\, ,\quad &
  (3\Delta+\Omega_{\rm{R}})/2\, ,
\end{eqnarray}
and the semiclassical dressed states
\begin{eqnarray}
  \label{eq:HamiltonianA3b}
  |\mu\rangle&=&{\cal{N}}_\mu
  (-\Omega |1\rangle+\Omega' |2\rangle)\, ,\nonumber\\
  |\phi\rangle&=&{\cal{N}}_\phi
  (2\Omega'|1\rangle+2\Omega |2\rangle+(\Delta-\Omega_{\rm{R}})|3\rangle)
  \, ,\nonumber\\
  |\psi\rangle&=&{\cal{N}}_\psi 
  (2\Omega'|1\rangle+2\Omega |2\rangle+(\Delta+\Omega_{\rm{R}})|3\rangle)\, ,  
\end{eqnarray}
where $\Omega_{\rm{R}}=(\Delta^2+4\Omega^2+4\Omega^{\prime 2})^{1/2}$,
${\cal{N}}_\mu=(\Omega^2+\Omega^{\prime 2})^{-1/2}$, 
${\cal{N}}_\phi=(2\Omega_{\rm{R}}(\Omega_{\rm{R}}-\Delta))^{-1/2}$ and
${\cal{N}}_\psi=(2\Omega_{\rm{R}}(\Omega_{\rm{R}}+\Delta))^{-1/2}$.
Now, using the new basis 
$\{ |0\rangle, |\mu\rangle, |\phi\rangle, |\psi\rangle\}$, 
we express the Hamiltonian~(\ref{eq:Hamiltonian1}) as
\begin{eqnarray}
  \label{eq:HamiltonianA4}
  H&=&\Delta\sigma_{\mu\mu}+(3\Delta-\Omega_{\rm{R}})/2\sigma_{\phi\phi}+
  (3\Delta+\Omega_{\rm{R}})/2\sigma_{\psi\psi}+\big({\cal{N}}_\mu g\Omega' a^{\dagger} \sigma_{0\mu}
  -2{\cal{N}}_\phi g\Omega a^{\dagger} \sigma_{0\phi}
  +2{\cal{N}}_\psi g\Omega a^{\dagger} \sigma_{0\psi}\nonumber\\
  &&-{\cal{N}}_\mu g'\Omega b^{\dagger} \sigma_{0\mu}
  -2{\cal{N}}_\phi g'\Omega' b^{\dagger} \sigma_{0\phi}
  +2{\cal{N}}_\psi g'\Omega' b^{\dagger} \sigma_{0\psi}+ {\rm{h.c.}}\big)\, .
\end{eqnarray}
Now we can eliminate atomic excited states $|\mu\rangle$, $|\phi\rangle$
and $|\psi\rangle$. To this end, we introduce a unitary
transformation~\cite{alexanianPRA95}
\begin{eqnarray}
  \label{eq:HamiltonianA5}
  U&=&\exp(S)\, ,
\end{eqnarray}
where
\begin{eqnarray}
  \label{eq:HamiltonianA6}
  S&=&\lambda_1(\sigma_{\mu 0} a -a^{\dagger} \sigma_{0\mu})
  +\lambda_2(\sigma_{\phi 0} a -a^{\dagger} \sigma_{0\phi})\nonumber\\
  &&+\lambda_3(\sigma_{\psi 0} a -a^{\dagger} \sigma_{0\psi})
  +\lambda_4 (b^{\dagger} \sigma_{0\mu}-\sigma_{\mu 0} b)+\lambda_5(\sigma_{\phi 0} b -b^{\dagger} \sigma_{0\phi})
  +\lambda_6(\sigma_{\psi 0} b -b^{\dagger} \sigma_{0\psi})
\end{eqnarray}
and $\lambda_i$ are dimensionless parameters such that 
$\lambda_i\sqrt{\langle a^{\dagger}a\rangle}$ and 
$\lambda_i\sqrt{\langle b^{\dagger}b\rangle}$ are very small compared
to $1$. These parameters will play the role of expansion parameters
associated with respective excited states. For example, $\lambda_1$
and $\lambda_4$ are associated with the state $|\mu\rangle$
(see equation~(\ref{eq:HamiltonianA6})).

We transform each operator in~(\ref{eq:HamiltonianA4}) using
the Baker–Hausdorf lemma
\begin{eqnarray}
  \label{eq:HamiltonianA7}
  X'&=&e^S X e^{-S}= X+[S,X]+(1/2!)\big[S,[S,X]\big]+\dots
\end{eqnarray}
If we choose $\lambda_1={\cal{N}}_\mu g\Omega'/\Delta$,
$\lambda_2=4{\cal{N}}_\phi g\Omega/(3\Delta-\Omega_{\rm{R}})$,
$\lambda_3=4{\cal{N}}_\psi g\Omega/(3\Delta+\Omega_{\rm{R}})$,
$\lambda_4={\cal{N}}_\mu g'\Omega/\Delta$,
$\lambda_5=4{\cal{N}}_\phi g'\Omega'/(3\Delta-\Omega_{\rm{R}})$,
$\lambda_6=4{\cal{N}}_\psi g'\Omega'/(3\Delta+\Omega_{\rm{R}})$
then terms which are linear in the field operators vanish in the
transformed Hamiltonian. If we moreover drop all terms much smaller than
$\lambda_1^2\Delta$ then we obtain the effective Hamiltonian~(\ref{eq:HamiltonianEFF}).
Note that both methods, \emph{i.e.} Reiter-S{\o}rensen method
and Alexanian-Bose method, give exactly the same formula for the effective 
Hamiltonian, despite the fact that both are just approximations.

It is also worth to note that the parameters $\lambda_i$ are given by
the ratios of the effective coupling constants to the dressed state energies.
The dressed state energies play the role of detunings in this dressed-state approach. 
So, $\lambda_i\ll 1$ means that the corresponding excited state is
very far off resonance from the ground atomic state $|0\rangle$, and thus,
its population is small. For instance, the smaller $\lambda_1$ and $\lambda_4$ are,
the smaller is the population of the state $|\mu\rangle$. Knowing this
we can obtain the conditions~(\ref{eq:condA}).

%\subsection*{Interaction with an external field}
\vspace*{5mm}
\noindent
\textbf{Interaction with an external field.} The two cavity modes interact according to the effective
Hamiltonian~(\ref{eq:HamiltonianEFF2}). The photon emission from the mode,
represented by $b$, into the waveguide is described by the Lindblad operator
$L_{\eta}=\sqrt{\eta}b$. The absorption in the mirrors for this mode
is modelled as the photon emission into an inaccessible mode and described by
$L_{\eta'}=\sqrt{\eta'}b$. The losses in the mirrors for the $a$ mode
are taken into account in the same manner. The photon absorption 
from the $a$ mode is described by the operator $L_{\kappa}=\sqrt{\kappa}a$.
Although the simplified model does not include atomic variables, spontaneous
emissions from the excited atomic states $|1\rangle$ and $|2\rangle$ are taken 
into account by assuming that there are two inaccessible travelling modes,
into which photons from both modes can be emitted in the way
described by the Lindblad operators $L_{\rm{eff}}^{(1)}$ and $L_{\rm{eff}}^{(2)}$.
The device working in the open mode has to transfer the state of the $a$ mode
to the waveguide. In order to calculate a quantity, which measures how close the
output field into waveguide is to the initial $a$ mode field, it is necessary
to describe the interaction of the quantum system with the accessible travelling mode. 
To this end, we use the input-output theory~\cite{collettPRA84,WallsMillburnBook,
tufarelliNJP12}, because it is perfectly suitable for the scheme illustrated in
figure~\ref{fig:systemEFF}. We have followed the treatment of Ref.~[\citenum{WallsMillburnBook}]
to derive the Heisenberg-Langevin equations for this scheme. These equations take the form
\begin{eqnarray}
  \label{eq:HLequations}
  \dot{a}&=&\big(-i\delta_2+(\sqrt{\zeta_1\theta_1}+\sqrt{\zeta_2\theta_2})/2\big) b+(\kappa+\zeta_1+\zeta_2) a/2
  -\sqrt{\kappa}a_{\kappa}
  -\sqrt{\zeta_1}c-\sqrt{\zeta_2}d\, ,\nonumber\\
  \dot{b}&=&\big(-i\delta_2+(\sqrt{\zeta_1\theta_1}+\sqrt{\zeta_2\theta_2})/2\big) a+\big(-i\delta_1+(\eta_{\rm{tot}}+\theta_1+\theta_2)/2\big)b-\sqrt{\eta}b_{\eta}-\sqrt{\eta'}b_{\eta'}-\sqrt{\theta_1}c-\sqrt{\theta_2}d\, ,
\end{eqnarray}
where $a_{\kappa}(t)$, $b_{\eta'}(t)$, $c(t)$ and $d(t)$ are output field 
operators of inaccessible travelling modes, $b_{\eta}(t)$ is the output field
operator of the waveguide mode and
$\zeta_1=n\gamma'\alpha_3^2 g^2$, $\theta_1=n\gamma'\alpha_1^2 g^{\prime 2}$,
$\zeta_2=n\gamma\alpha_2^2 g^2$, $\theta_2=n\gamma\alpha_3^2 g^{\prime 2}$,
$\eta_{\rm{tot}}=\eta'+\eta$. The matrix form of equations~(\ref{eq:HLequations})
is given by
\begin{eqnarray}
  \label{eq:HLequationsVec}
  \dot{\boldsymbol{v}}&=&\boldsymbol{M} \boldsymbol{v}-\boldsymbol{v_{\rm{out}}}\, ,
\end{eqnarray}
with
\begin{equation}
  \label{eq:M}
\boldsymbol{M}\equiv\begin{bmatrix}
\frac{\kappa+\zeta_1+\zeta_2}{2}
&\frac{\sqrt{\zeta_1\theta_1}+\sqrt{\zeta_2\theta_2}}{2}-i\delta_2\\
\frac{\sqrt{\zeta_1\theta_1}+\sqrt{\zeta_2\theta_2}}{2}-i\delta_2
&\frac{\eta_{\rm{tot}}+\theta_1+\theta_2}{2}-i\delta_1
\end{bmatrix}\, ,
\end{equation}
where $\boldsymbol{v}=[a, b]^{\rm{T}}$ and $\boldsymbol{v_{\rm{out}}}=
[\sqrt{\kappa}a_{\kappa}+\sqrt{\zeta_1}c+\sqrt{\zeta_2}d,
\sqrt{\eta}b_{\eta}+\sqrt{\eta'}b_{\eta'}+\sqrt{\theta_1}c+\sqrt{\theta_2}d]^{\rm{T}}$.

%\subsection*{Figure of merit}
\vspace*{5mm}
\noindent
\textbf{Figure of merit.}
Now, we can follow closely the treatment of Tufarelli~\emph{et al.}~\cite{tufarelliPRL14}
to get the figure of merit of the scheme. First, we have to define the bosonic operator
for the waveguide field travelling away from the device 
\begin{eqnarray}
  \label{eq:rel4}
  f_{\rm{out}}&\equiv&\int_0^\infty u(\tau) b_{\eta}(\tau) d\tau\, ,
\end{eqnarray}
with $u(\tau)$ being a temporal profile of the form
\begin{eqnarray}
  \label{eq:rel5}
  u(\tau)&\equiv& \frac{[e^{-\boldsymbol{M}\tau}]_{1,2}}
  {\sqrt{\int_0^\infty |[e^{-\boldsymbol{M}\tau}]_{1,2}|^2 d\tau}}\, .
\end{eqnarray}
Next, we introduce the bosonic operator $h_{\rm{ext}}$ representing all
inaccessible travelling modes. We do not need to know the specific form
of $h_{\rm{ext}}$ in our calculations. Then we can relate the annihilation
operator $a$ at the time $t=0$ to the output modes using
the formula
\begin{eqnarray}
  \label{eq:rel6}
  a(0)&=&\sqrt{F} f_{\rm{out}}-\sqrt{1-F} h_{\rm{ext}}\, ,
\end{eqnarray}
where
\begin{eqnarray}
  \label{eq:fig_mer}
  F&=&\eta\int_0^\infty \big|[e^{-\boldsymbol{M}\tau}]_{1,2}\big|^2 d\tau\, .
\end{eqnarray}
It is worth to note the similarity between equation~(\ref{eq:rel6}) and
a unitary transformation representing a beam splitter of transmittance $F$.
This similarity allows us to consider an abstract beam splitter described by
relations
\begin{eqnarray}
  \label{eq:beamsplitter}
  a(0)&=&\sqrt{F} f_{\rm{out}}-\sqrt{1-F} h_{\rm{ext}}\, ,\\
  a_{\rm{vac}}(0)&=&\sqrt{1-F} f_{\rm{out}}+\sqrt{F} h_{\rm{ext}}\, .
\end{eqnarray}
The abstract mode $a_{\rm{vac}}(0)$ must be empty, because the total
excitation number has to be conserved, \emph{i.e.}, the initial number of photons
inside the $a$ mode has to be equal to the total number of photons inside 
outgoing modes $f_{\rm{out}}$ and $h_{\rm{ext}}$.
Using the abstract beam-splitter model of the device it is easy to get
formula for $f_{\rm{out}}$:
\begin{eqnarray}
  \label{eq:formula}
  f_{\rm{out}}&=&\sqrt{F} a(0)+\sqrt{1-F} a_{\rm{vac}}(0)\, .
\end{eqnarray}
The parameter $F$ satisfies $0\leq F\leq 1$ and,
as it is easy to see from equation~(\ref{eq:formula}), it can work
as a figure of merit, because as $F$ gets closer to one, the output field
$f_{\rm{out}}$ gets closer to the initial field $a(0)$.
This fact is especially clearly seen in the Schr\"odinger picture~\cite{tufarelliPRL14}
\begin{eqnarray}
  \label{eq:rho}
  \rho_{\rm{out}}&=&e^{(1-F){\cal{L}}}\rho_0\, ,
\end{eqnarray}
where $\rho_0$ is the initial state of the $a$ mode, $\rho_{\rm{out}}$
is the final state of the $f_{\rm{out}}$ mode and the Liouvillian is given by
\begin{eqnarray}
  \label{eq:L}
  {\cal{L}}\rho&=&\frac{1}{2}\big(2 a\rho a^{\dagger}-a^{\dagger}a\rho-\rho a^{\dagger}a\big)\, .
\end{eqnarray}
In order to investigate how well the initial quantum state can be
extracted from the cavity using the device presented in figure~\ref{fig:systemEFF},
we have to express the figure of merit $F$ as a function of parameters of
this device. It can be done using the method presented in Ref.~[\citenum{tufarelliPRL14}].
First, we express the figure of merit as
\begin{eqnarray}
  \label{eq:fig_mer2}
  F&=&\eta {\cal{X}}_{1,2,2,1}(\boldsymbol{M})\, ,
\end{eqnarray}
where
\begin{eqnarray}
  \label{eq:XM1221}
{\cal{X}}_{1,2,2,1}(\boldsymbol{M})&=&
\int_0^\infty [e^{-\boldsymbol{M}\tau}]_{1,2} [e^{-\boldsymbol{M}^{\dagger}\tau}]_{2,1} d\tau  
\end{eqnarray}  
is an element of the tensor ${\cal{X}}$. We can express this tensor in the
matrix form as
\begin{eqnarray}
  \label{eq:XM}
{\cal{X}}(\boldsymbol{M})&=&
\int_0^\infty e^{-\boldsymbol{M}\tau}\otimes e^{-\boldsymbol{M}^{\dagger}\tau} d\tau\, ,  
\end{eqnarray}
where $\otimes$ indicates the Kronecker product. Since ${\cal{X}}(\boldsymbol{M})$ is the
solution to a Sylvester equation, we can obtain all elements of ${\cal{X}}(\boldsymbol{M})$
just by solving linear system of equations
\begin{eqnarray}
  \label{eq:XMlinear}
(\boldsymbol{M}\otimes I)\,{\cal{X}}(\boldsymbol{M})
+{\cal{X}}(\boldsymbol{M})\,(I\otimes \boldsymbol{M}^{\dagger})&=&I\otimes I\, ,
\end{eqnarray}
where $I$ indicates the $2\times 2$ identity matrix. In this way we derive the
formula for ${\cal{X}}_{1,2,2,1}(\boldsymbol{M})$, which we insert into
equation~(\ref{eq:fig_mer2}). Unfortunately, the obtained expression is too
complex to be useful, and thus, it is necessary to resort to further
approximations. If we assume that 
$\eta_{\rm{tot}}\gg\delta_2\gg\kappa\, , \zeta_1\, , \theta_1\, , \zeta_2\, , \theta_2$
then the figure of merit can be well approximated by
\begin{eqnarray}
  \label{eq:Ffinalb} 
  F=\frac{\eta}{\eta_{\rm{tot}}}\Big[1
-\frac{(\sqrt{\zeta_1\theta_1}+\sqrt{\zeta_2\theta_2})^2}{2\delta_2^2}-\frac{\eta_{\rm{tot}}\big(\kappa+\zeta_1+\zeta_2\big)}
{4\delta_2^2+\eta_{\rm{tot}}\big(\kappa+\zeta_1+\zeta_2\big)}\Big]\, .
\end{eqnarray} 
%\end{document}

%------------------------------------------------------------------
% \bibliography{wneka}
% \bibliographystyle{naturemag}

%------------------------------------------------------------------
\section*{Acknowledgements}

K.B. acknowledges the support by the Polish National Science Centre (Grant No.
DEC-2013/11/D/ST2/02638).

\section*{Author Contributions}

G.C. developed the theoretical framework and performed calculations. 
All authors discussed the methods, analysed the results, and participated in the manuscript preparation.

\section*{Additional information}

\noindent\textbf{Competing Interests:} The authors declare that
they have no competing interests.

\noindent\textbf{Correspondence and requests for materials} should
be addressed to G.C.~(email: chimczak@amu.edu.pl).

\end{document}